\documentclass[reprint,superscriptaddress,amsmath,amssymb,aps,prlshowkeys,acknowledgments]{revtex4-1}

\usepackage{graphicx}
\usepackage{dcolumn}
\usepackage{bm}
\usepackage{hyperref}

\usepackage{color}
\usepackage[normalem]{ulem}

\newcommand{\lm}[1]{\color{black}{#1}}
\newcommand{\new}[1]{\color{black}{#1}}

\begin{document}

\title{Machine learning based localization and classification with atomic magnetometers}

\author{Cameron Deans}
\affiliation{Department of Physics and Astronomy, University College London, Gower Street, London WC1E 6BT, United Kingdom}

\author{Lewis D. Griffin}
\affiliation{Department of Computer Science, University College London, Gower Street, London WC1E 6EA, United Kingdom}

\author{Luca Marmugi}
 \email{l.marmugi@ucl.ac.uk}
\affiliation{Department of Physics and Astronomy, University College London, Gower Street, London WC1E 6BT, United Kingdom}

\author{Ferruccio Renzoni}
\affiliation{Department of Physics and Astronomy, University College London, Gower Street, London WC1E 6BT, United Kingdom}

\date{\today}

\begin{abstract}
We demonstrate identification of position, material, orientation and shape of  objects imaged by an $^{85}$Rb atomic magnetometer performing electromagnetic induction imaging supported by machine learning. Machine learning maximizes the information extracted from the images created by the magnetometer, demonstrating the use of hidden data. Localization 2.6 times better than the spatial resolution of the imaging system and successful classification up to 97$\%$ are obtained. This circumvents the need of solving the inverse problem, and demonstrates the extension of machine learning to diffusive systems such as low-frequency electrodynamics in media. Automated collection of  task-relevant information from quantum-based electromagnetic imaging will have a relevant impact from biomedicine to security.
\end{abstract}


\maketitle

\onecolumngrid
\begin{center}
This is a preprint version of the article appeared in Physical Review Letters:\\
C. Deans, L. D. Griffin, L. Marmugi, F. Renzoni, Phys. Rev. Lett. {\bf 120}, 033204 (2018). \\DOI: \href{https://doi.org/10.1103/PhysRevLett.120.033204}{10.1103/PhysRevLett.120.033204}\\
\end{center}
\vskip 15pt
\twocolumngrid

Electromagnetic induction imaging (EMI), or magnetic induction tomography (MIT), with atomic magnetometers (AMs) was recently demonstrated for mapping the electric conductivity of objects and imaging of metallic samples \cite{apl, ol, arne, opex}.

EMI and its classical counterpart with conventional magnetic field sensors \cite{mitgriffiths} rely on the detection of the AC magnetic field generated by eddy currents excited in media. This poses severe problems for image reconstruction, particularly in cluttered contexts or with low-conductivity specimens. The inherently diffusive and non-linear nature of low-frequency electrodynamics in media makes conventional ray-optics analysis impossible. Consequently, back-projection approaches  \cite{backprojection} are of limited use. Furthermore, the solution of the inverse problem for low-frequency electromagnetics is ill-posed, undetermined, and computationally challenging \cite{merwa2005}. Ultimately, these limitations reduce the attainable information from EMI and its spatial resolution. 

In this Letter, we propose and demonstrate machine learning (ML) \cite{mlreview} as a method for enhancing the EMI capabilities and circumventing the problem of image reconstruction and interpretation. ML has thus far been applied in a wealth of fields \cite{functionals, polymers, glass, plosonera, mlgenomics, entanglement, loops, infinite, mlquantum}. ML-aided security screening in the X band \cite{lewis1,lewis2} and biomedical imaging have been widely demonstrated \cite{parkinsonsimaging, opticaltomography}, as well as image reconstruction through scattering media in the optical band \cite{scatteringmedia, lensless}. All these applications to well-established imaging technologies are underpinned by linear systems, with ray-like propagation.

Here we present proof-of-concept demonstrations of  EMI by an AM with metallic and non-metallic samples. Their localization, and material, orientation and shape classification from low-resolution images is aided by ML. AM-EMI supported by ML maximizes the information obtained from the images and provides relevant data for specific tasks, without requiring the inverse problem. This improves or enables identification of critical features, such as structural defects in non-destructive evaluation \cite{apl, arne}, concealed threats in screening applications \cite{opex}, or conductivity anomalies in biological tissues \cite{scirep}.  New perspectives open up for high performance imaging based on EMI, in particular with AMs, with a relevant impact on science and society.

\begin{figure*}[!htbp]
\includegraphics[width=\linewidth]{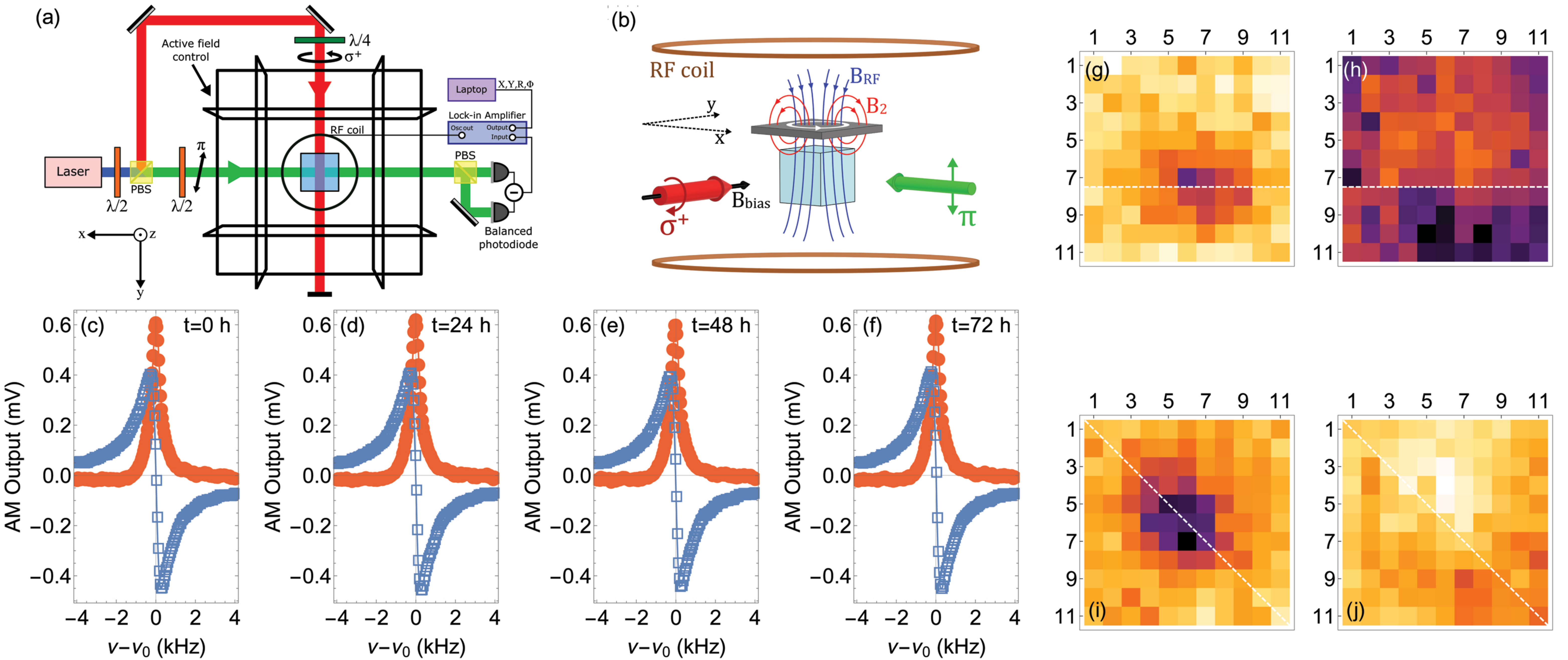}
\caption{\textbf{(a)} {\lm Sketch of the experimental apparatus. PBS indicates a polarizing beam splitter. \textbf{(b)} Details of the imaging technique. Samples are moved by an XY stage in the plane marked by dotted arrows. B$_{RF}$ is the RF field driving the AM and inducing eddy currents (in white). The EMI secondary field is drawn in red. \textbf{(c)-(f)} Frequency response of the RF AM, in operational conditions, at $\nu_{0}$=200~kHz: in-phase (X, filled circles) and quadrature (Y, empty squares) components. Traces were acquired in the same conditions every 24~h. \textbf{(g)-(j)} Examples of low-resolution EMI images used for this work: Al rectangle (50$\times$25$\times$3~mm$^{3}$, $\sigma$=3.77$\times$10$^{7}$~S/m), obtained with 8~mm scan step size at 200~kHz. \textbf{(g)} Amplitude (R) image at 0$^{\circ}$ orientation. \textbf{(h)} Corresponding phase ($\Phi$) image. \textbf{(i)}~R image at +45$^{\circ}$ orientation. \textbf{(j)} Corresponding $\Phi$ image. Raw data displayed, with the same color scale for R and $\Phi$.  A thin dashed line marks the orientation of the rectangle main axis.}}\label{fig:overview}
\end{figure*}

In our setup, imaging relies on the position-resolved detection of the secondary magnetic field produced by eddy currents induced in the sample of interest by an AC magnetic field (B$_{RF}$), oscillating at $\nu_{RF}$=200~kHz with amplitude 1.9$\times$10$^{-8}$~T. This primary field is produced by a pair of 180~mm diameter Helmholtz coils. The imaged object's response is detected by an $^{85}$Rb radio-frequency optically pumped atomic magnetometer (AM), coherently driven by the same field (B$_{RF}$){\lm, and described in \cite{apl, opex}. We recall that a $\sigma^{+}$-polarized beam resonant with the F=3$\rightarrow$F'=4 transition of the $^{85}$Rb D$_{2}$ line aligns the atomic spins via optical pumping. B$_{RF}$ excites a time-varying transverse component of the atomic polarization, which is perturbed by the secondary field produced by the objects of interest. The effects of this are imprinted in the polarization plane rotation of a $\pi$-polarized laser probe (Faraday rotation), detuned by +420~MHz. Further details can be found in the Supplementary Material \cite{supplementary}.}

Samples are moved with respect to the AM by a computer-controlled XY stage in the (x,y) plane  (Fig.~\ref{fig:overview}(a)), 30~mm above the AM and enclosed by the RF Helmholtz coils. Four images per scan are obtained by simultaneously measuring the in-phase (X) and quadrature (Y) components of the AM output, as well as its amplitude (R) and phase-lag ($\Phi$).  Active environmental and stray fields control \cite{opex} ensure continuous and consistent operation for more than 240 consecutive hours in an unshielded environment, without requiring any human intervention. To maintain consistency over long measurement timescales, the AM sensitivity is purposely decreased to 3.3$\times$10$^{-11}$~T/$\sqrt{Hz}$ \cite{supplementary}, but it is  stable throughout the measurement campaigns (Figs.~\ref{fig:overview}(c)-(f)). 

Each image is an 11$\times$11 (22$\times$22) matrix, where each point corresponds to an 8~mm (4~mm) step of the XY stage. The center-position of the sample is randomly distributed within the interval $[$0, 30$]\times[$0, 30$]$~mm$^{2}$, with a resolution of 10$^{-3}$~mm. The large RF coils and the long XY stage step severely reduced the images resolution and contrast (see also \cite{supplementary}), beyond the point where the samples and their details can be directly identified from the images{\lm. As an example, we present images of an Al rectangle obtained at 200~kHz with 8~mm scan step size in Figs.~\ref{fig:overview}(g)-(j).}  We show that ML compensates for such degradation, shifting the burden of image recognition from the observer to the computer, and from the imaging phase to the training process.

We have tested the following combined tasks: i) Classification of four materials and localization; ii)  Classification of four orientations and localization; and iii) Classification of four shapes and localization. In each case, 255 sets of images of a same class (e.g. same material) were collected, with random position. Each set comprises four images, namely X, Y, R, and $\Phi$.  In total, 4080 datafiles per task were available for training. Sets of 160 \textit{blind} datafiles were acquired to test more realistic conditions: the blind images were acquired in different measurement runs, without any direct correlation to the training datasets. Details of these images were hidden during ML analysis.

ML algorithms were tested, tuned and validated using random splits of the images, grouped into disjoint testing and training subsets (cross-validation) \cite{mlstandard, new}. The latter had an arbitrarily chosen size of N. The optimum parameters were then applied to the corresponding blind datasets. Deviations from the groundtruth (i.e. the set of nominally true values) were measured with the root-mean-square-error (RMSE) of the distance between the predicted position and the groundtruth positions for localization. Failure rate ($\varepsilon$) was introduced for classifications  \cite{supplementary}. RMSE and $\varepsilon$ were then investigated as a function of the number of training images N. Results were compared to baseline and ceiling performance. Baseline performance is the error obtained with the best blind guess.  For localization, this corresponds to RMSE$_{l}$=12.38~mm; for classifications to $\varepsilon_{l}$=75$\%$. The ceiling performance is the minimum error achievable given the characteristics of data and their variability. For localization, this corresponds to half the XY stage step (RMSE$_{h}$=4~mm, unless otherwise stated); for classifications, to  $\varepsilon_{h}$=0$\%$. The corresponding 95$\%$ confidence intervals (CI) were calculated via bootstrap resampling-with-replacement of the blind data, to avoid the assumption that errors were normally-distributed \cite{bootstrap}.

Among the available datasets, we found the $\{$R,$\Phi$$\}$ combination to be the most effective in the explored tasks \cite{supplementary}. It is therefore presented as the primary combination in the following analysis. Linear regressors (LR), nearest neighbors, random forest, neural networks and support vector machine (SVM) \cite{svm} algorithms were tested. {\lm Linear regression was found to uniformly perform better for the localization tasks. Radial Basis Function (RBF) Kernel SVM was the best} performing for the classification tasks \cite{supplementary}. RBF SVM parameters, namely margin softness and gaussian width, were tuned by cross-validation \cite{crossvalidation} within the training data.

For material classification and localization, four squares of equal size, made of copper, aluminium, Ti90/Al6/V4 alloy and graphite respectively, were tested. Their conductivities vary between 5.98$\times$10$^{7}$~S/m to 7.30$\times$10$^{4}$~S/m, making the graphite the lowest conductivity and the first non-metallic sample to be imaged with EMI performed by an AM \cite{apl, arne, opex, spiecameron}. 

Cross-validation of localization with LR performed better than baseline performance beyond only N=8 training images: RMSE=9.7~mm with CI=$[$7.2~mm, 12.3~mm$]$. The ceiling performance was achieved with N=64 training images. A final RMSE=3.1~mm with CI=$[$3.0~mm, 3.2~mm$]$ was obtained with N=2048. We underline that this residual error is 2.6 times smaller than the XY step size: this constraint is surpassed by ML.

Material classification with RBF SVM produced an error $\varepsilon$=2\% with N=2048 training images, and of only 12\% with  N=8 \cite{supplementary}. This result is aided by the different EMI signals of the four materials: the amplitude and the phase-lag of the secondary field are proportional to the specimen's conductivity \cite{mitgriffiths}. The difference in the EMI images demonstrates the capability of our system of discriminating different materials. However, it also makes material identification - even in non-optimum conditions - potentially achievable by a human operator (see also \cite{supplementary}).

We therefore focus on more challenging tasks in the following. In these cases, a blind dataset is also collected at a later stage and tested with the ML algorithms tuned during training.

Task ii) comprises varying position and orientation. In this case,  Al rectangles with aspect ratio 2:1 were aligned in four different orientations, namely $\{-45^{\circ}, 0^{\circ}, +45^{\circ}, +90^{\circ}\}$ with respect the system's quantization axis ($\hat{y}$). Two rectangles were tested, sample A ($50\times25\times3$~mm$^{3}$) and sample B ($40\times20\times3$~mm$^{3}$). For A, the XY stage step size was maintained at 8~mm. For B, two settings at 8~mm and 4~mm were used. Figure~\ref{fig:orientation} shows the results in terms of classification errors $\varepsilon$.

\begin{figure}[htbp]
\includegraphics[width=\linewidth]{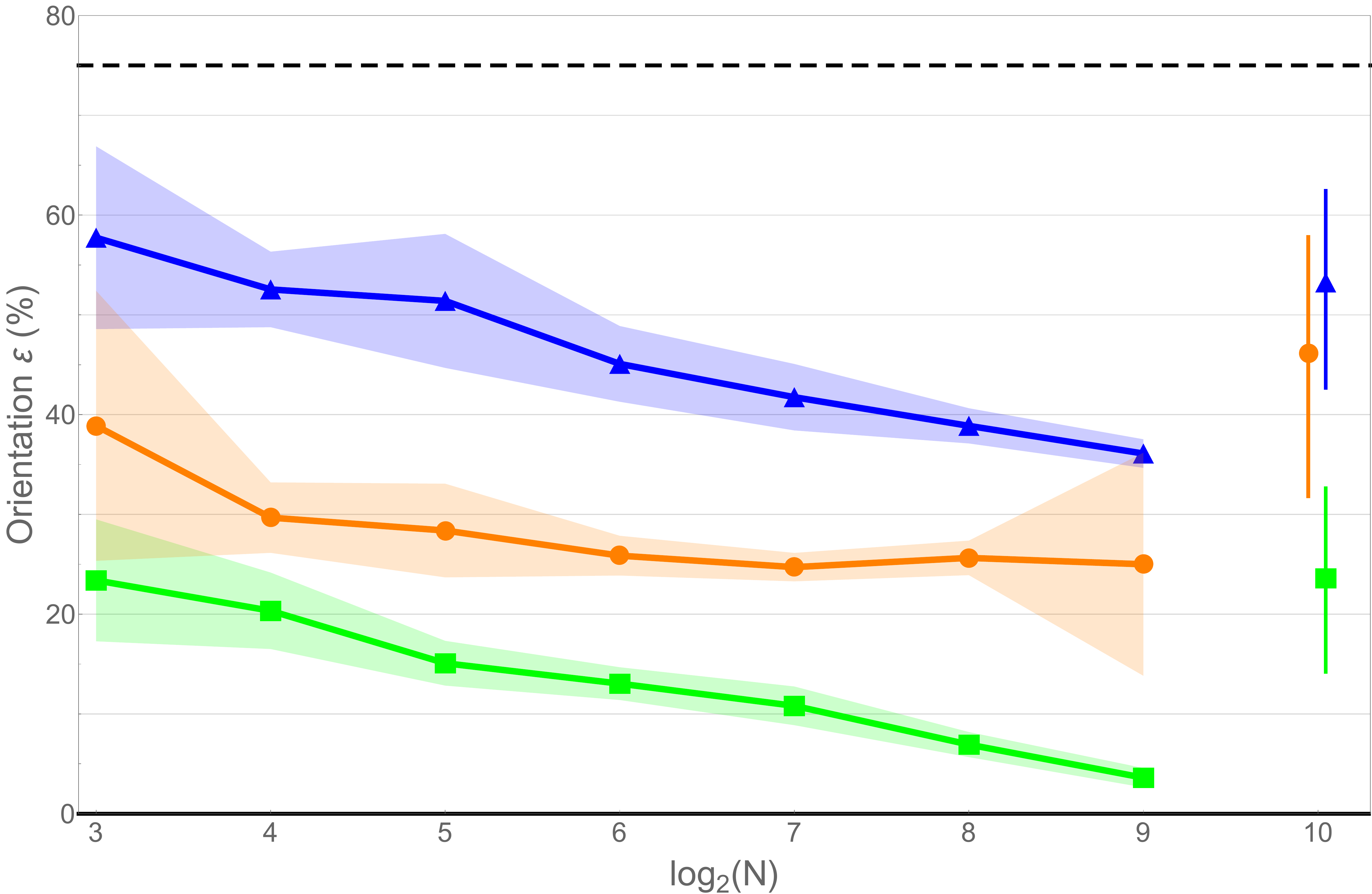}
\caption{Orientation $\varepsilon$ versus N. Green squares: sample A, 8~mm step. Blue upwards triangles: sample B, 8~mm step. Orange disks: sample B, 4~mm step. Isolated markers are the blind dataset results. The shaded area marks the 95\% CI of the mean cross-validated performance. The error bars for the blinded results indicate the 95\% CIs of system performance given the limited amount of blinded data. The  dashed line marks the baseline performance ($\varepsilon_{l}$=75$\%$). The thick horizontal line marks the ceiling performance ($\varepsilon_{h}$=0$\%$).}\label{fig:orientation}
\end{figure}

Cross-validation was successfully applied in each case. A consistent improvement in performance is observed for increases in training samples and reduction in step size. Sample A error rapidly converges towards  $\varepsilon \leq$20\%, and reaches $\varepsilon$=4\% with N=512. This performance is obtained also with a N=40 blind dataset, where we have obtained a residual  $\varepsilon$=20\%, with CI=$[8\%,32\%]$. We note that such a task, even in the most favorable case of sample A, would be virtually impossible for a human observer (see Figs.~\ref{fig:overview}(g)-(j)).

A similar behavior for cross-validation is observed also in the worst case of sample B at 8~mm step size: $\varepsilon < \varepsilon_{l}=75\%$ with N=8 training images. However, overall weaker performance of sample B is observed. This is related to the size of the test object: sample B is 1.56 times smaller than sample A. Consequently, the level of EMI signals obtained is lower, thus reducing the images' contrast. Statistical fluctuations explain the larger CI of the blind B dataset with 4 mm steps: this set has only N=20 acquisition. {\lm Such small random variations in blind datasets were identified by the ML algorithms only. Nevertheless, classification via ML is not hampered:} the blind results for both samples are well below the random choice level. Furthermore, results are improved by reducing the XY step size. The scan density impact is highlighted in Fig.~\ref{fig:orientation}: the two-times shorter step produces a reduction between 50\% and 25\% of $\varepsilon$.

Interestingly, the different orientations and the different $\varepsilon$ do not affect the localization prediction (as shown in Fig.~\ref{fig:position}). Even when the blind classification is challenged as in the case of sample B, the localization RMSE becomes smaller than the ceiling performance with only N=256 (8~mm dataset). The trend is confirmed by the analysis of the 4~mm step size dataset with sample B (RMSE$_{l}$=2~mm, in this case). A RMSE=4~mm is obtained with N=512, and - as a further confirmation of statistical fluctuations in blind images - the blind data set performed better than the cross-validation sets (RMSE=2.6~mm, CI=$[$1.9~mm, 3.2~mm$]$).

\begin{figure}[htbp]
\includegraphics[width=\linewidth]{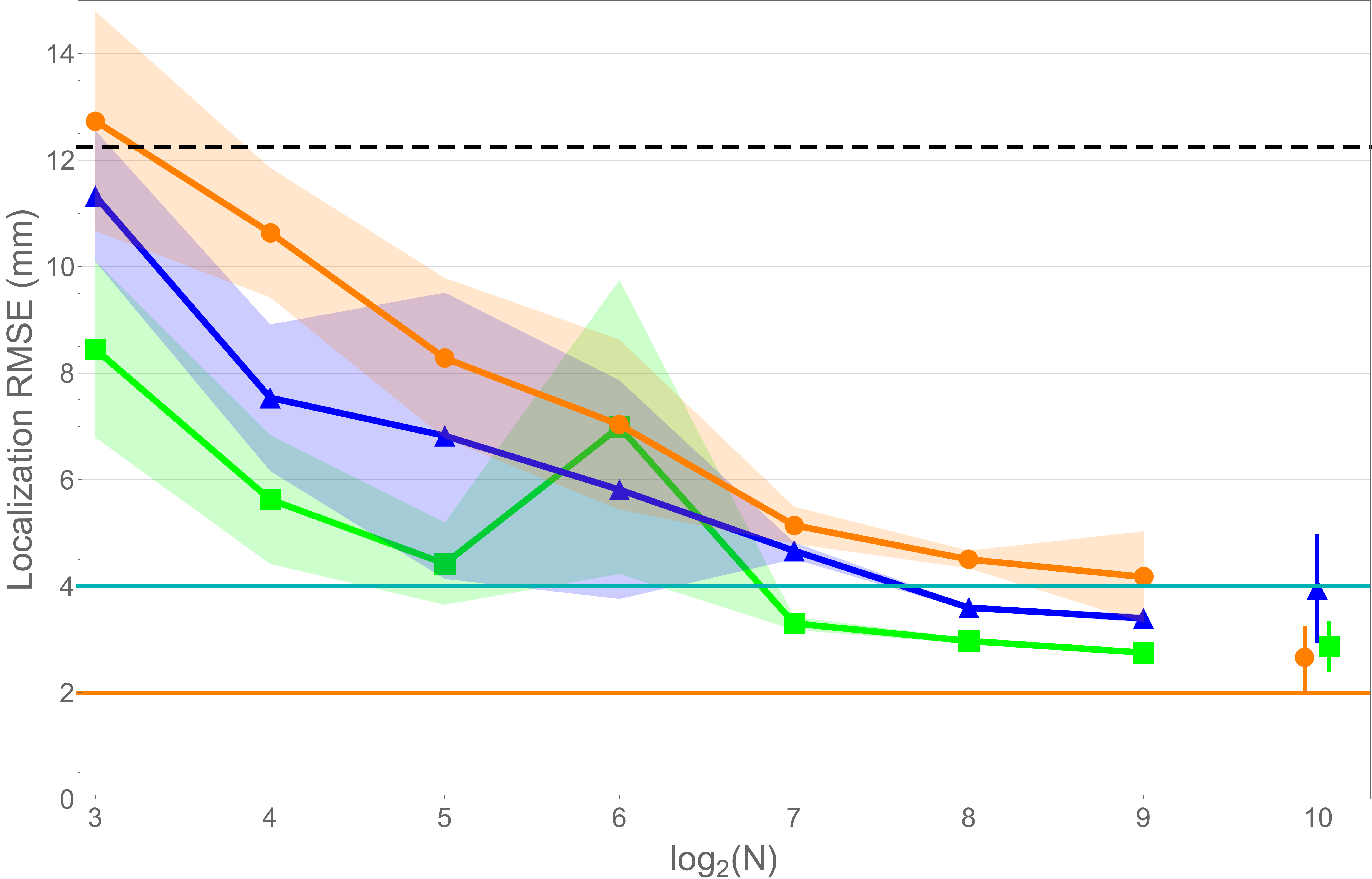}
\caption{Localization RMSE versus N. Green squares: sample A, 8~mm step. Blue upwards triangles: sample B, 8~mm step. Orange disks: sample B, 4~mm step. Isolated markers are the blind dataset results. The shaded area marks the 95\% CI of the mean cross-validated performance. The error bars for the blinded results indicate the 95\% CIs of system performance given the limited amount of blinded data. The dashed line marks the baseline performance (RSME$_{l}$=12.38~mm). The thick horizontal lines mark the ceiling performance RMSE$_{h}$=4~mm and 2~mm, respectively.}\label{fig:position}
\end{figure}

For the last task four shapes of Al, 3~mm thick, were used: a 50~mm square, a 50~mm disk, a 50~mm$\times$25~mm rectangle, and an isosceles triangle, base 50~mm and height 50~mm. 

Localization performs very well, aligned with previous results, confirming its substantial independence from other degrees of freedom.  We obtained RMSE=4.8~mm, CI=$[$4.6~mm, 5.0~mm$]$ with N=512. The N=40 blind dataset gave a consistent RMSE=4.5~mm, CI=$[$3.6~mm, 5.4~mm$]$ \cite{supplementary}.

Shape classification results are shown in Fig.~\ref{fig:shapes}. Classification outperforms chance with only N=8 images. The best  performance, $\varepsilon$=3\%, CI=$\left[2\%,4\%\right]$ is obtained with N=512. Consistent performance is observed with the N=40 blind dataset:  $\varepsilon$=22\%,  CI=$[$9$\%$,35$\%]$.

\begin{figure}[htbp]
\includegraphics[width=\linewidth]{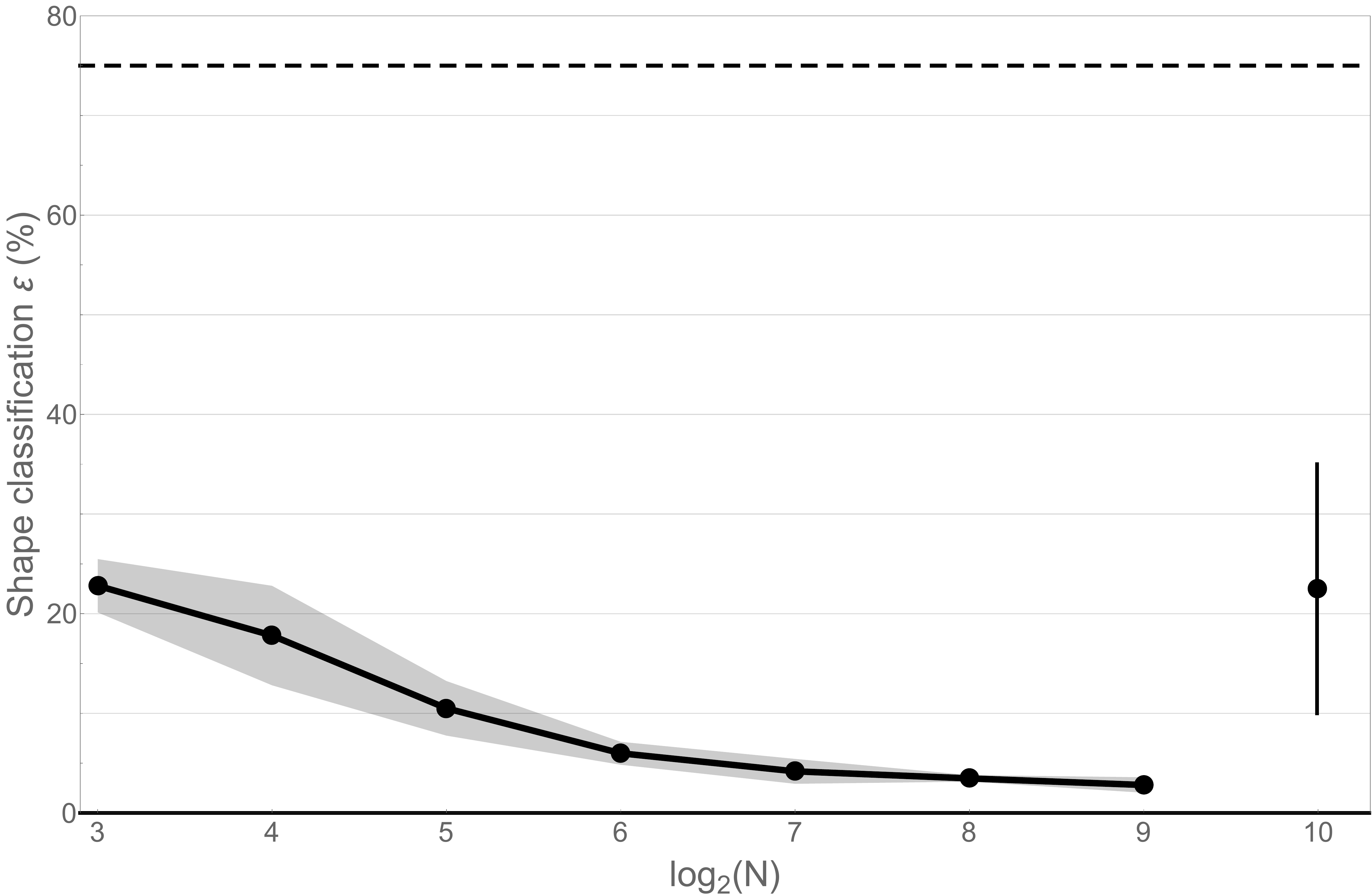}
\caption{Shape classification $\varepsilon$ versus N. The isolated mark is the blind dataset result. The shaded area marks the 95\% confidence interval of the mean cross-validated performance. The error bars for the blinded results indicate the 95\% CIs of system performance given the limited amount of blinded data. The dashed line marks the baseline performance ($\varepsilon_{l}$=75$\%$). The thick horizontal line marks the ceiling performance ($\varepsilon_{h}$=0$\%$).}\label{fig:shapes}
\end{figure}

Virtually no incorrect attributions were observed between the square and the disk, whereas around 6\% symmetric confusion was observed between the rectangle and the triangle (Tab.~\ref{tab:confusionshapes}). This is attributed to the equal surface (1250~mm$^{2}$) of the two samples. Given that the RF Helmholtz coils' diameter is about three times larger than the samples' cross-section, eddy currents are excited within their entire surface. With the four samples having the same thickness, the surface area becomes relevant for discrimination.

\begin{table}[htbp]
\begin{center}
\begin{tabular}{c|c|c|c|c|}
{} & Square & Disk & Rectangle & Triangle \\
\hline
\textit{Square} &  \textbf{99.5\%} & 0.5\% & 0.0\% & 0.0\% \\
\hline
\textit{Disk} & 0.2\%  & \textbf{99.8\%}  & 0.0\% &  0.0\%\\
\hline
\textit{Rectangle} & 0.0\% & 0.2\% & \textbf{93.6\%} & 6.2\% \\
\hline
\textit{Triangle} & 0.0\% & 0.0\% & 6.7\%  & \textbf{93.3\%}  \\
\hline
\end{tabular}
\caption{Shape classification confusion matrix based on N=256. Rows are predictions.} \label{tab:confusionshapes}
\end{center}
\end{table}

In light of this, we note that shape classification is successful even with same areas. This clearly shows that the ML is capable of implicitly using and integrating all the information, including those hidden in the samples' area and in other parameters when areas are not effective, to improve the classification.

In conclusion, we have demonstrated that EMI with AMs provides information for identifying position, material, orientation and shape of test samples, thanks to ML support. ML was applied for the first time to diffusive, non-ray-optics images produced by an $^{85}$Rb RF AM operating in EMI modality. {\lm This demonstrates the suitability of ML for diffusive and complex physical systems}. Localization better than the smallest pixel of the image was demonstrated. Classification by ML allowed the identification of low-conductivity materials such as graphite, imaged for the first time with an AM. ML has also revealed the use of hidden information such as the deduced area of the samples, without any specific input. {\lm Based on the present results, no evidence preventing scaling to larger number of classes was found, provided that the chosen imaging resolution matches the scale of the relevant features.}

Our {\lm findings} demonstrate that the conventional approach for EMI/MIT image reconstruction (the electromagnetic inverse problem) can be efficiently circumvented by ML, with relevant impact on computational burden. This opens up new perspectives for EMI/MIT with atomic magnetometers.

\begin{acknowledgments}
\vskip -2pt
{\lm This work was funded by the UK Quantum Technology Hub in Sensing and Metrology, Engineering and Physical Sciences Research Council (EPSRC) (EP/M013294/1).} C.~D. acknowledges support from the Engineering and Physical Sciences Research Council (EPSRC) (EP/L015242/1).\end{acknowledgments}

\pagebreak
\newpage

\onecolumngrid
\begin{center}
{\textbf{\large Supplemental Material for ``Machine Learning Based Localization and Classification with Atomic Magnetometers''}}\label{supp}
\vskip 15pt

%
%
\end{center}

\date{\today}


\maketitle

Supporting material for ``Machine Learning Based Localization and Classification with Atomic Magnetometers'' by C.~Deans \textit{et al.} is presented here.

\section{$^{85}$Rb Radio-Frequency Atomic Magnetometer for Electromagnetic Induction Imaging}\label{sec:am}
The hardware of the sensor is based on an $^{85}$Rb radio-frequency atomic magnetometer (RF AM), with active stabilization of the background stray fields  \cite{opex}, thus capable of reliable operation in unshielded environments for several days. Rubidium atoms are contained in a cubic quartz cell (side 25~mm), filled with 20~Torr of N$_{2}$ to reduce atomic diffusion. Atoms are spin-polarized by $\sigma^{+}$ pump beam (P$_{pump}$=650~$\mu$W) resonant to the D$_{2}$ F=3$\rightarrow$F$^{'}$=4 transition, with a collinear DC magnetic field $|\mathbf{B}_{bias}|$=4.26$\times$10$^{-5}$~T. This produces optical pumping to the $|F=3, m_{F}=+3\rangle$ state. 

Atomic coherences are excited and driven by an AC magnetic field (B$_{RF}$), along the vertical axis, of amplitude 1.9$\times$10$^{-8}$~T. A $\pi$ polarized probe beam (P$_{probe}$=50~$\mu$W), detuned by +400~MHz with respect to the center of the F=3$\rightarrow$F$^{'}$=4 transition, intercepts the atoms perpendicularly to the pump beam, and maps their Larmor precession with its polarization plane rotation (Faraday rotation). The probe beam is analyzed by a polarimeter, whose output is selectively amplified by a computer-controlled lock-in amplifier (LIA).

Aiming at the highest degree of stability and fidelity over several days, the operation frequency is fixed at $\nu_{RF}$=200~kHz, the vapor cell is maintained at 298~K. In these conditions, we recorded an RF sensitivity \cite{savukov2005, chalupczak2012} of $\delta$B$_{RF}$=3.3$\times$10$^{-11}$~T/$\sqrt{Hz}$, continuously maintained without any intervention and measured without averaging or integration. In the same unshielded arrangement and without averaging, we have measured a maximum sensitivity of 3$\times$10$^{-12}$~T/$\sqrt{Hz}$, in optimized conditions.

The RF driving field is produced by a pair of 180~mm diameter circular Helmholtz coils, supplied by the LIA internal oscillator. This geometry ensures greater homogeneity and a finer control of the AM's drive, with a positive impact on the AM's performance and a further 10-fold reduction of the required RF field amplitude with respect to the best conditions of recent works \cite{opex}. The electromagnetic induction imaging (EMI)/magnetic induction tomography (MIT) primary field frequency $\nu_{RF}$ can be easily tuned to match the best requirements for each sample \cite{spiecameron}, or to penetrate thick barriers \cite{opex}. However, in the present case, it was kept fixed to obtain unbiased datasets and allow the machine learning (ML) to take into account the samples' different responses. This allowed: obtaining truly \textit{blind} datasets; increasing of the imaging speed; and avoiding any human intervention.

The broader distribution of the EMI primary field causes a degradation of the EMI imaging, compared to previous results with smaller coils \cite{apl, opex}, as discussed in the following.

\section{Imaging Performance}\label{sec:imaging}
\begin{figure}[htbp]
\includegraphics[width=0.7\linewidth]{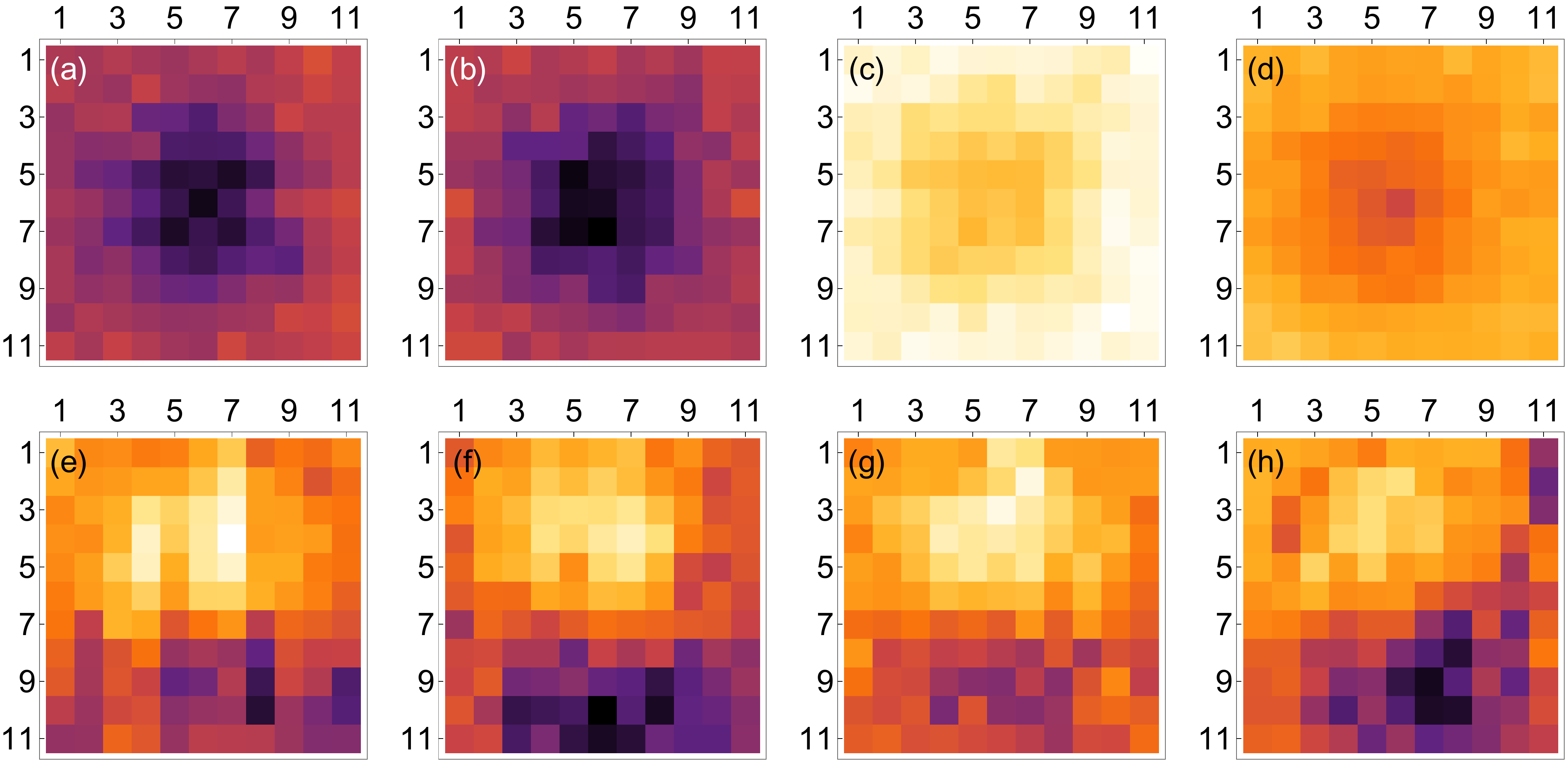}
\caption{Examples of images at $\nu_{0}$=200~kHz and step size 8~mm, used in the main text.  \textbf{(a)}, \textbf{(e)} R and $\Phi$, respectively, of the Cu square. \textbf{(b)}, \textbf{(f)} R and $\Phi$, respectively, of the Al square. \textbf{(c)}, \textbf{(g)} R and $\Phi$, respectively, of the Ti90/Al6/V4 square. \textbf{(d)}, \textbf{(h)} R and $\Phi$, respectively, of the graphite square. Raw data plotted against the same color scale for R and $\Phi$.} \label{fig:examples}
\end{figure}

In Fig.~\ref{fig:examples}, examples of R and $\Phi$ images of different materials obtained in this work are shown. 

We have tested Cu, Al, Ti90/Al6/V4 alloy and graphite. The latter is - to date - the lowest conductivity sample imaged with AM operating in EMI modality, and the first non-metallic one. We have imaged the graphite sample's in-plane conductivity $\sigma_{\parallel}$=7.30$\times$10$^{4}$~S/m \cite{graphite}. An overview of the materials' relevant characteristics is shown in Tab.~\ref{tab:materials}, as well as their skin depths at $\nu_{RF}$=200~kHz. We also introduce the effective thickness $\ell$, {\lm defined as the ratio between the sample's thickness $t$ and the material's skin depth $\delta$}:

\begin{equation}
\ell(\nu) = \dfrac{t}{\delta (\nu)}~.
\end{equation}

\begin{table}[h]
\begin{center}
\begin{tabular}{|c|c|c|c|}
\hline
\textbf{Material} & \textbf{Electric Conductivity} & \textbf{Skin Depth (200~kHz)} & \textbf{$\ell$ (200~kHz)} \\
\hline
Copper &  5.98$\times$10$^{7}$~S/m \cite{metals} & 0.146~mm & 13.7 \\
\hline
Aluminium & 3.77$\times$10$^{7}$~S/m \cite{metals} & 0.183~mm & 11.0 \\
\hline
Ti90/Al6/V4  &  5.95$\times$10$^{5}$~S/m \cite{tialv} & 1.63~mm & 0.6 \\
\hline
Graphite  &  7.30$\times$10$^{4}$~S/m ($\sigma_{\parallel}$) \cite{graphite} & 3.61~mm & 1.4 \\
\hline
\end{tabular}
\caption{Overview of the materials tested.}\label{tab:materials}
\end{center}
\end{table}

\begin{figure}[htbp]
\begin{center}
\includegraphics[width=0.30\linewidth]{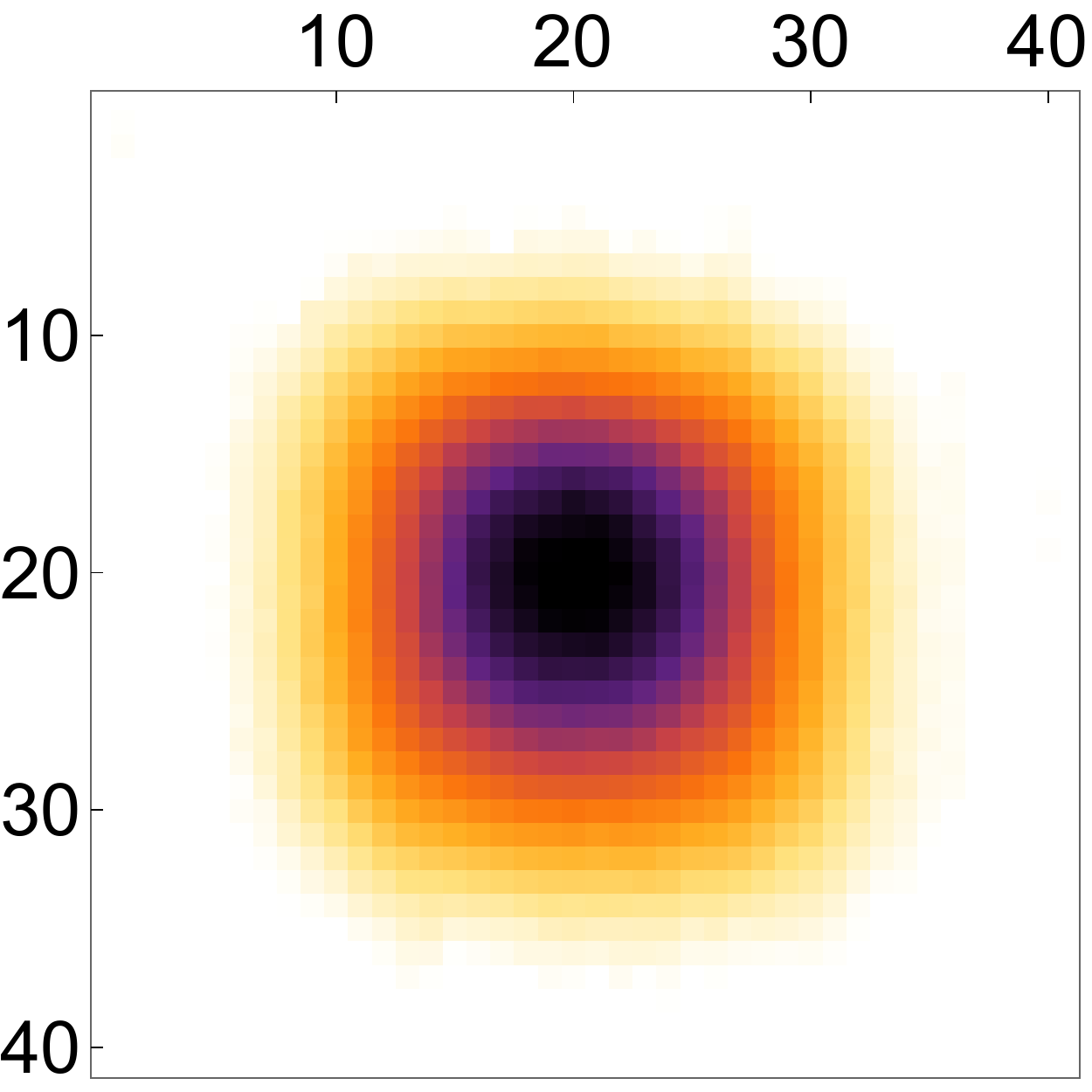}
\caption{Graphite square (side 50~mm, thickness 5~mm, X component), imaged with an RF atomic magnetometer in optimum conditions.  $\nu_{RF}$=150~kHz supplied by a 7.8~mm diameter ferrite core coil and 2~mm scan step size. Raw data.}\label{fig:graphite}
\end{center}
\end{figure}

The low spatial resolution imposed by the step size (the 8~mm step size implies that - on average - only a few measurement points match the surface of the sample) and by the broad RF field  (B$_{RF}$) clearly degrade the image quality and contrast. The broad distribution of B$_{RF}$, more than three times larger than the largest side of the tested specimens, results in the simultaneous excitation of eddy currents over the entire surface of the sample, further complicated by their diffusion. This implies that - at a given position - the AM will always detect a contribution from the whole of the object. In Fig.~\ref{fig:graphite}, we show an example of the impact of the B$_{RF}$ spatial distribution across the sample: an image of the same graphite square, obtained in optimum conditions at 150~kHz, and with 2~mm XY step size and 7.8~mm diameter ferrite core coil, is presented.

\section{Data Analysis}\label{sec:data}
For all datasets, images were arranged in 2$\times$11$^{2}$=242-dimensional vectors with random splits of data of size N. This constitutes the feature space for ML. {\lm In other words, N is the size of the dataset used for ML}. Algorithms were tuned and validated on these datasets (cross-validation), before analyzing the blind datasets.

Various combinations of $\{$X$\}$, $\{$Y$\}$, $\{$R$\}$, and $\{\Phi\}$ (feature sets) data were tested, and their performance evaluated for the specific task, namely localization or classification of materials, orientations, and shapes. Different algorithms were tested to identify the best approach.

\begin{figure}[htbp]
\includegraphics[width=0.49\linewidth]{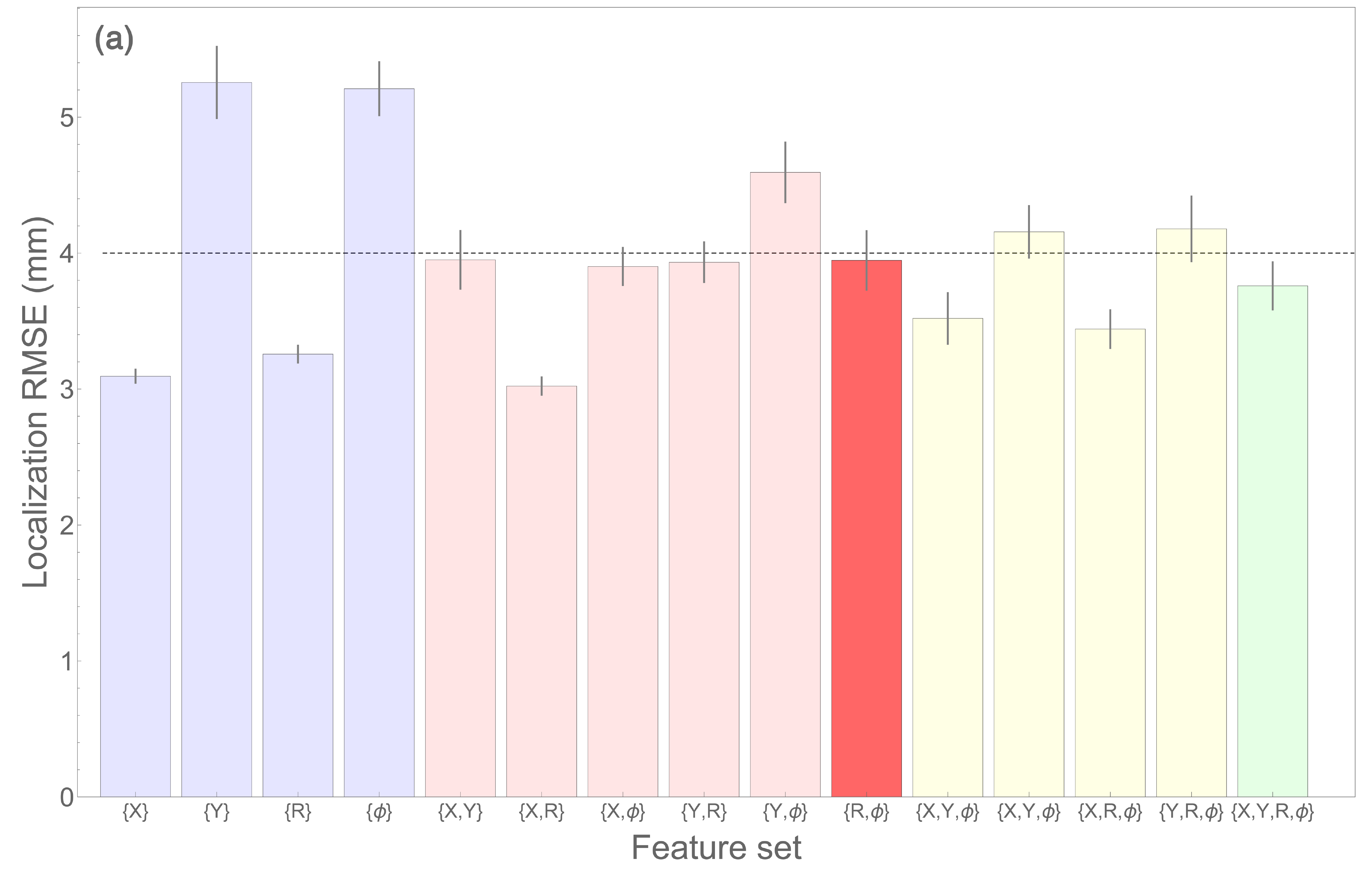}
\includegraphics[width=0.49\linewidth]{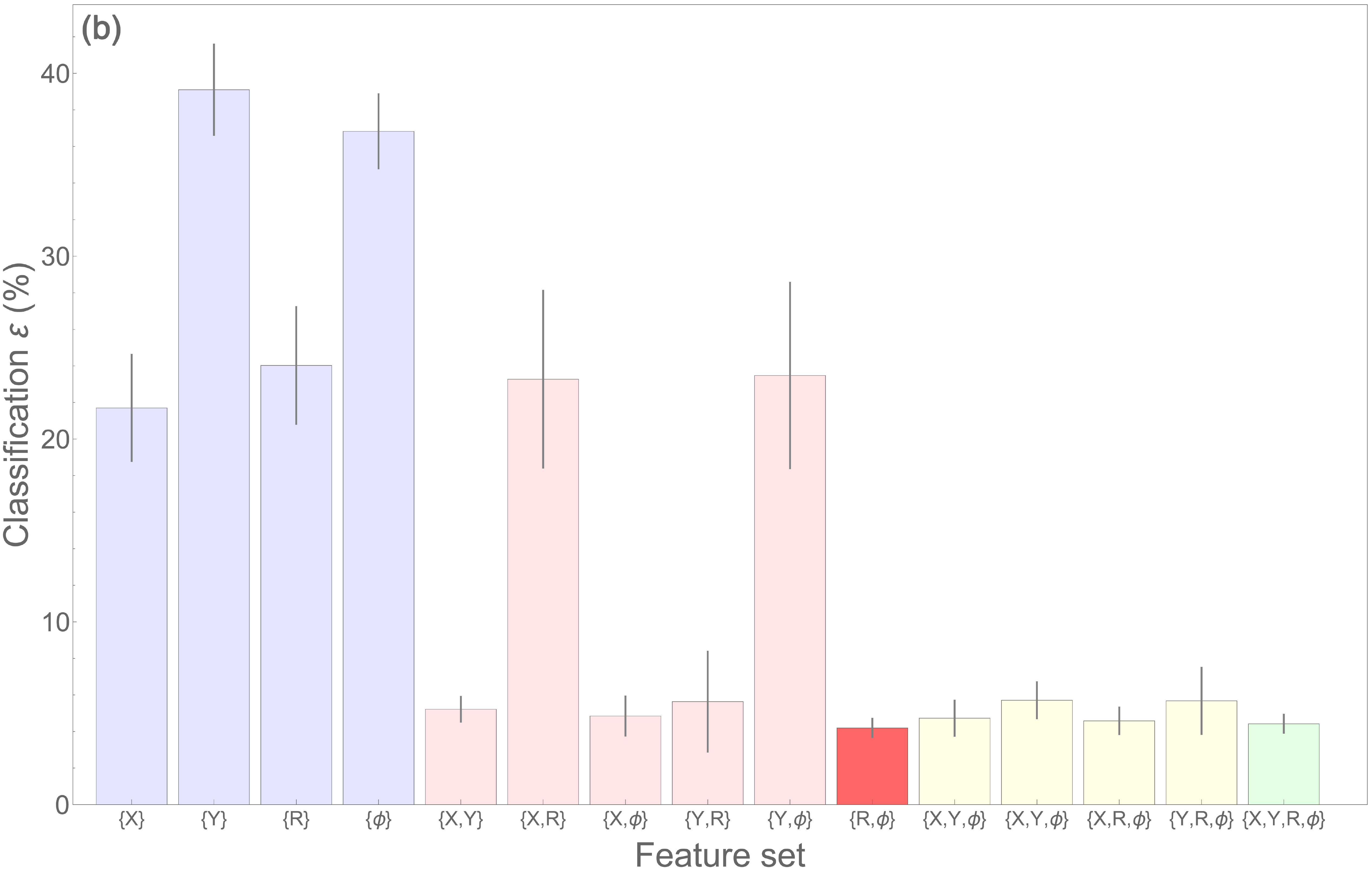}
\caption{Identification  of the optimum feature set for ML: performance of different feature sets for the materials dataset. \textbf{(a)} Localization RMSE. The dotted line marks the ceiling performance, corresponding to half the step of XY stage. \textbf{(b)} Material classification error. Note that  $\{$X+R$\}$ indicates concatenated feature vectors, not added values.}\label{fig:dataset}
\end{figure}

For localization, shown in Fig.~\ref{fig:dataset}(a) in the case of materials datasets, all the combinations of feature sets performed much better than the baseline (chance) performance, and in many cases better than the ceiling performance RMSE$_{h}$=4~mm. Generally speaking, the feature sets $\{$X$\}$ and $\{$R$\}$ allowed for smaller RMSEs than $\{$Y$\}$ and $\{ \Phi \}$. We note that combination of feature sets were obtained by concatenating the corresponding vectors.

A similar behavior was observed in the case of material classification (Fig.~\ref{fig:dataset}(b)). All the combinations of feature sets performed much better than the baseline (chance) level (75\%), but more dramatic differences are found between different feature sets. For example, $\{$R+$\Phi\}$ reduced the classification error $\varepsilon$ to a minimum, with an 8-fold improvement with respect to $\{$Y$\}$ alone. We attribute this to the larger sensitivity of $\Phi$ to the conductivity of the target \cite{opex} and, consequently, to the material it is made of.

Furthermore, we recall that - as a consequence of the LIA operation - only two pairs of datasets are truly independent: X and R, and Y and $\Phi$ are related to each other. This explains the comparable performance of related parameters and the minimal improvement seen for their combinations (e.g. $\{$X+R$\}$ compared to $\{$X$\}$ or $\{$R$\}$ alone). In contrast, significant improvements over single feature sets are seen when combining non-related parameters (eg. $\{$X+Y$\}$, compared to $\{$X$\}$ or $\{$Y$\}$ alone). However, generally speaking, more features do not necessarily imply better performance, even when independent: although they potentially provide more information to exploit, more features also require higher dimensional feature spaces. This substantially increases the amount of training and the overall workload. Accordingly, we chose the best feature set (i.e. $\{$R+$\Phi\}$) as that providing the best performance, with the smallest number of features.

\begin{figure}[htbp]
\includegraphics[width=0.35\linewidth]{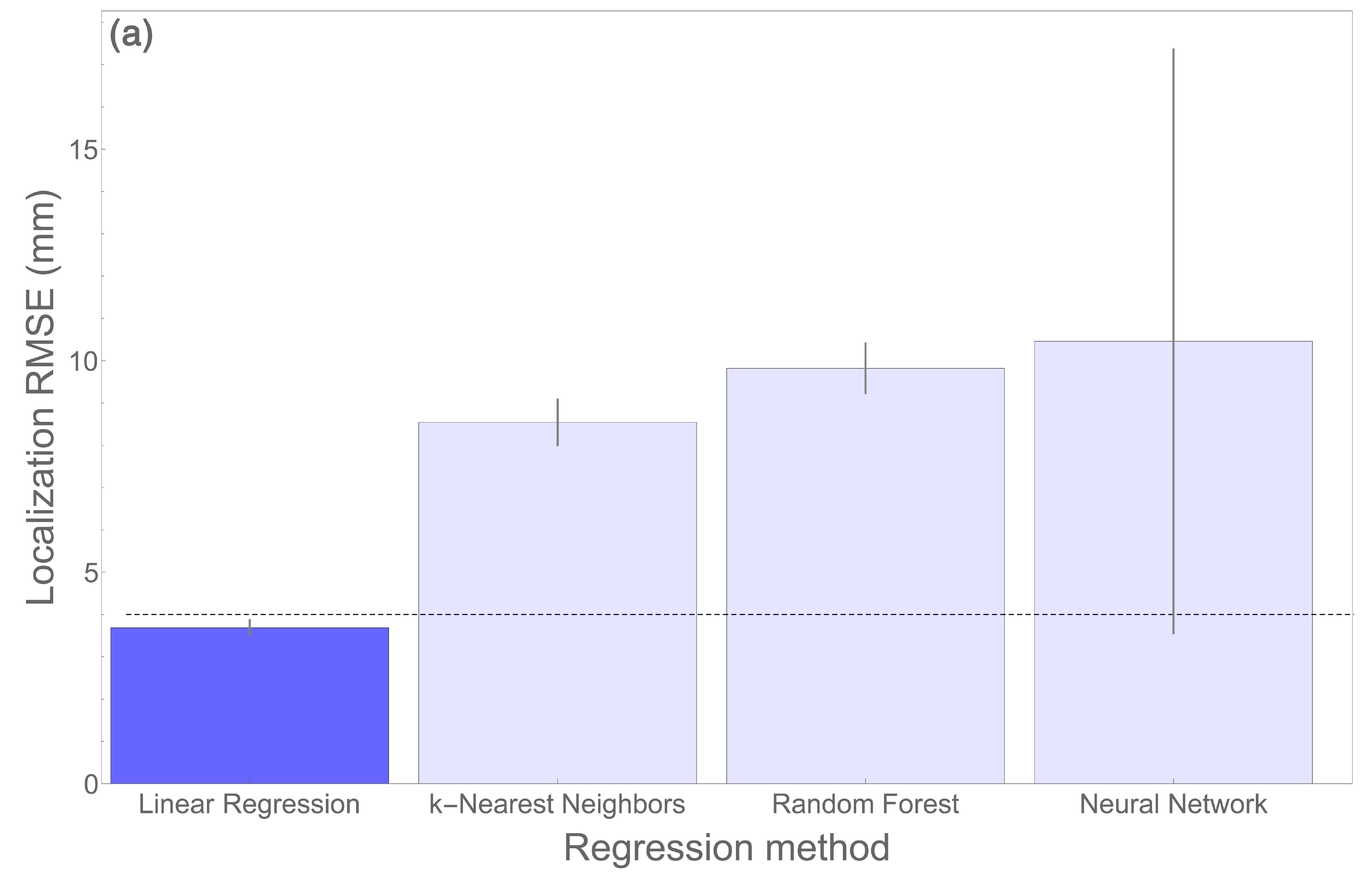}
\includegraphics[width=0.35\linewidth]{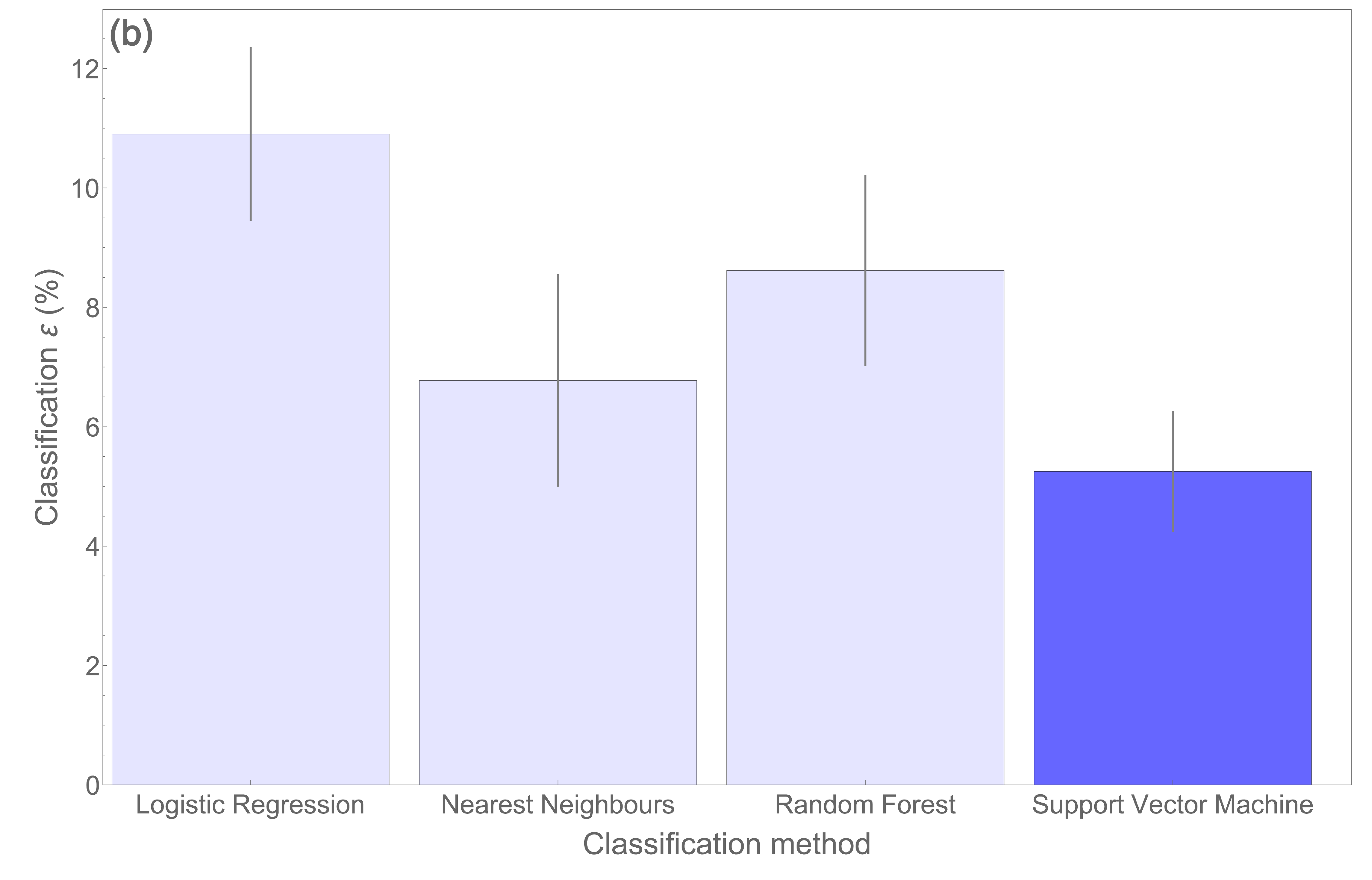}
\includegraphics[width=0.35\linewidth]{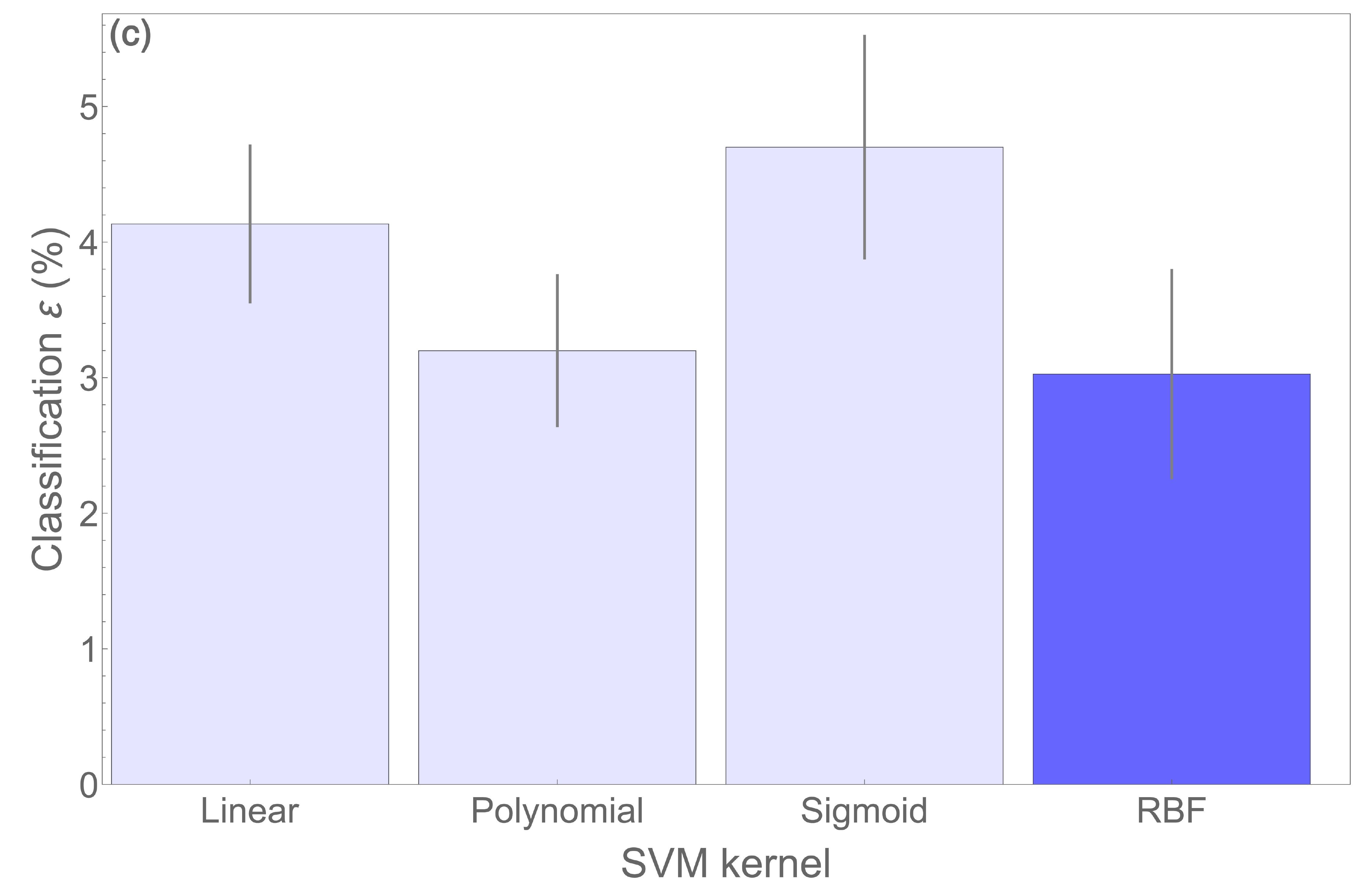}
\caption{{\lm Identification  of the optimum algorithm for ML: performance of different algorithms for the material and shape dataset. \textbf{(a)} Localization RMSE. The dotted line marks the ceiling performance, corresponding to half the step of XY stage.  \textbf{(b)} {\new Material classification error. In both cases k=3 was optimised by cross validation.} \textbf{(c)} Shape classification error. In this case, best performing homogeneous polynomial kernel was found to be degree 2.}}\label{fig:algorithms}
\end{figure}

Comparative tests were also conducted with different algorithms for regression (i.e. localization) and classification. Final performance tests, after optimization on the training datasets, were compared. An example of such results is shown in Fig.~\ref{fig:algorithms}.

\begin{figure}[htbp]
\includegraphics[width=0.32\linewidth]{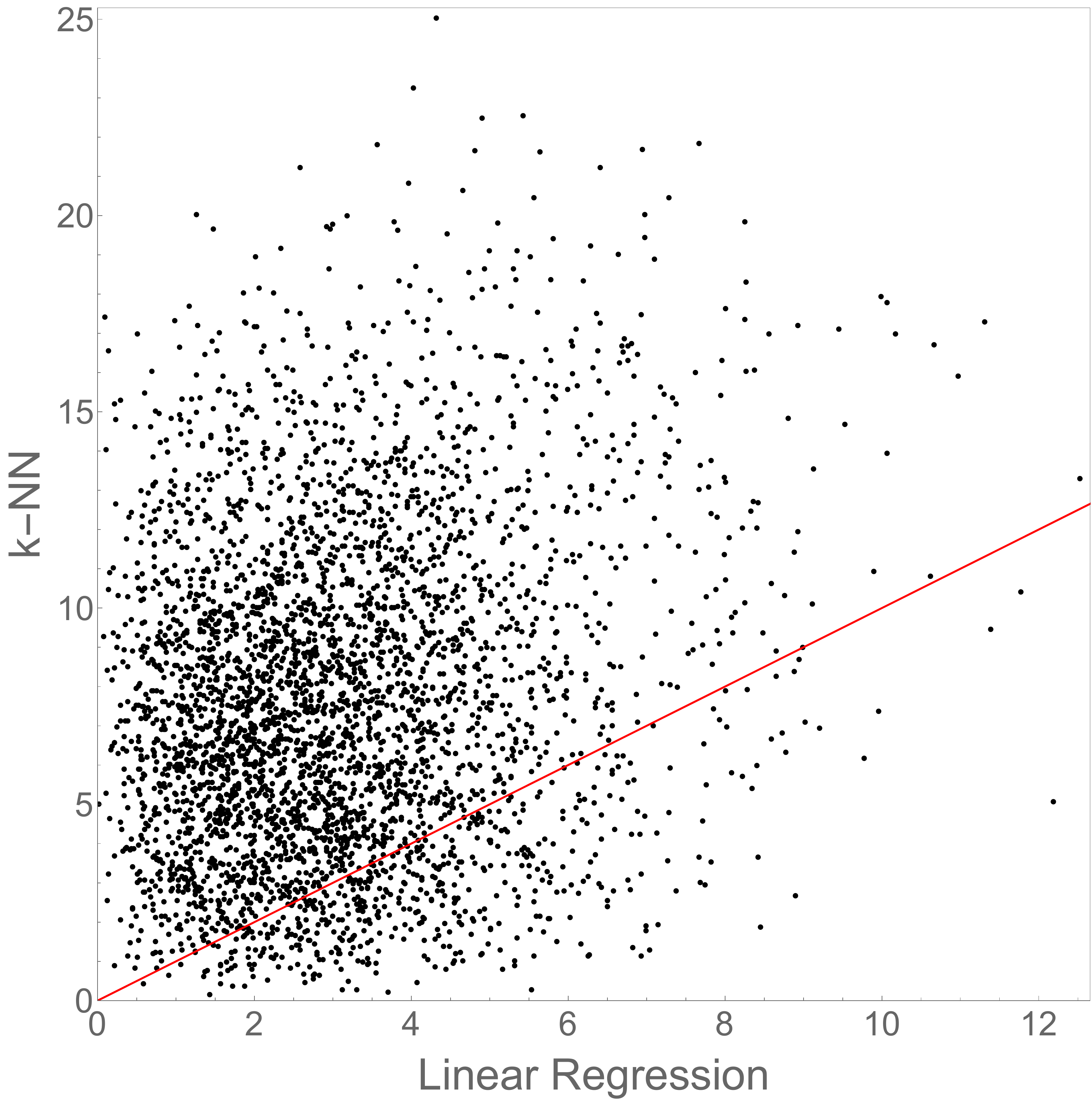}
\caption{{\lm Identification of the optimum algorithm for localization: scatter plot of RMSE obtained linear regression versus k-nearest neighbor. The continuous line marks the y=x line, indicating equal performance of the two algorithms.}}\label{fig:scatter}
\end{figure}

All localizations performed best with linear regression (LR), which surpassed ceiling performance (Fig.~\ref{fig:algorithms}(a)). The better performance of LR with position prediction is due to the strong prior of linear relationship among data underpinning LR. This is well-suited and completely exploited in the case of localization tasks, regardless of the other varying parameter (material, orientation or shape){\lm , and across the explored range of errors. This is demonstrated in Fig.~\ref{fig:scatter} in the case of k-nearest neighbors (k-NN), and it is confirmed by vectors error analysis: k-NN produces residuals errors more than two times larger in all directions, regardless the position of the test object. We also found that a neural network approach was unstable with our datasets. This explains the larger uncertainty in Fig.~\ref{fig:algorithms}(a)}.

Classifications performed best with tuned soft Support Vector Machine (SVM) with a Radial Basis Function (RBF) Kernel, as shown in Fig.~\ref{fig:algorithms}(b). SVM finds the hyperplane of the feature space, which divides the data in classes while maximizing the margin around the  hyperplane that contains no data. Here, we used a RBF Kernel, which allows non-linear boundaries to separate the data classes, by expressing the candidate boundaries in terms of weighted sums of gaussian functions centered on each datapoint. The width of the gaussian functions was controlled by a tunable parameter. {\new The RBF Kernel cross-validation tuning provided a length scale sigma corresponding to $\sim$70$\%$ of the total dispersion of the measurement values.} Some data are allowed to enter the margin separating the two hyperplanes (soft-margin SVM), with a tolerance controlled by the margin softness parameter. Both parameters were tuned during the cross-validation over the training datasets.

In the main text, orientation and material classifications were performed by using Mathematica's (11.1.1.0) inbuilt SVM to automatically tune hyper-parameters by cross-validation. SVMs were built between all pairs of class labels, and the overall label chosen by voting. Class priors were uniform.

For testing on blind data, regressors and classifiers were built from randomly chosen subsets of the training data of the indicated sizes. This process was repeated to allow standard deviations of performance to be computed, to estimate the performance consistency of the approach.

\subsection{Localization and classification figures of merit}
For each dataset with $n$ samples, therefore a total of $4n$ images, namely R, $\Phi$, X and Y, we have introduced the following sets:

\begin{enumerate} 
\item{$\mathcal{T}=\{t_{1}, t_{2},..., t_{n}\}$ of groundtruth values;}
\item{$\mathcal{P}=\{p_{1}, p_{2},..., p_{n}\}$, which are the {\lm experimental samplings (i.e. the outcome of the ML algorithm applied to experimental data)} of the groundtruth distribution  $\mathcal{T}$ belongs to.}
\end{enumerate}

\subsubsection{Root-Mean-Square Error}
To assess the ``goodness'' of the predictions, the root-mean-square-error (RMSE) is used, as standard practice. RMSE is defined as the expectation value of the squared difference between the predictions and groundtruth values:

\begin{equation}
RMSE=\sqrt{E((p_{i}-t_{i})^{2})}=\sqrt{\sum_{i=1}^{n} \dfrac{(p_{i}-t_{i})^{2}}{n}}~.\label{eqn:rmse}
\end{equation}

For localization, we have introduced the cartesian distance between $p_{i}$=(x$_{i}$,y$_{i}$) and $t_{i}$= (x$_{0,i}$,y$_{0,i}$):

\begin{equation}
d_{i}=\sqrt{(x_{i}-x_{0,i})^{2}+(y_{i}-y_{0,i})^{2}}~.
\end{equation}

In this view, the $\{ d_{i} \}$ represent the deviations, per each measurement, from the true position of object, and all the results can be compared to each other. The RMSE in this case simply becomes:

\begin{equation}
RMSE_{d}=\sum_{i=1}^{n} \dfrac{d_{i}}{\sqrt{n}}~.
\end{equation}

\subsubsection{Success Rate}
This figure of merit quantifies the ``goodness'' of classifications. By introducing a counter of successes $s_{i}$, the classification success rate $S$ is given by:

\begin{equation}
S = 100 \cdot \sum_{i=1}^{n} \dfrac{s_{i}}{n}~,\label{eqn:success}
\end{equation}

\noindent where the success event $s_{i}$ is defined as:

\begin{equation}
s_{i} \quad : \quad s_{i}=
\begin{cases}
1 \quad \text{for} \quad p_i=t_i,\\
0 \quad \text{for} \quad p_i\neq t_i~.
\end{cases} \label{eqn:s}
\end{equation}

From Eq.~\ref{eqn:s}, consistently with the RMSE analysis, we calculate the classification error $\varepsilon$ as:

\begin{equation}
\mathcal{\varepsilon}=100-S~.\label{eqn:error}
\end{equation}

\section{Datasets Details}\label{sec:details}
For each task described in the main text, the training dataset comprises 255 samples per classification category (i.e. 1020), each simultaneously imaged in R, $\Phi$, X and Y. This leads to a total 4080 files per training dataset. The position of the sample is randomly chosen by a computer algorithm in the interval $[ \Delta$x,$\Delta$y$]$=($[$0, 30$]\times[$0, 30$]$~mm$^{2}$), with a precision of 10$^{-3}$~mm. This ensures the absence of unwanted correlations among data and, in the long run, averaging of any possible transient or local effect.

For training, images belonging to a same class are sequentially acquired, before being analyzed only once the dataset acquisition is complete. Blind datasets comprise 40 samples (10 per class, total 160 files), and are acquired with the same approach, but in different measurements sessions, after the AM has been completely reset. This ensures the absence of unwanted correlations between blind and training datasets, thus mimicking realistic conditions, where the imaging campaign is well-separated from the training. The anonymized blind dataset then is analyzed with ML.

The numbers are reduced to 520 (2080 files) in the case of 4~mm resolution training, and 20 (80 files) for the blind dataset.

\subsection{Material dataset - Classification and localization}
\begin{figure}[htpb]
\includegraphics[width=0.49\linewidth]{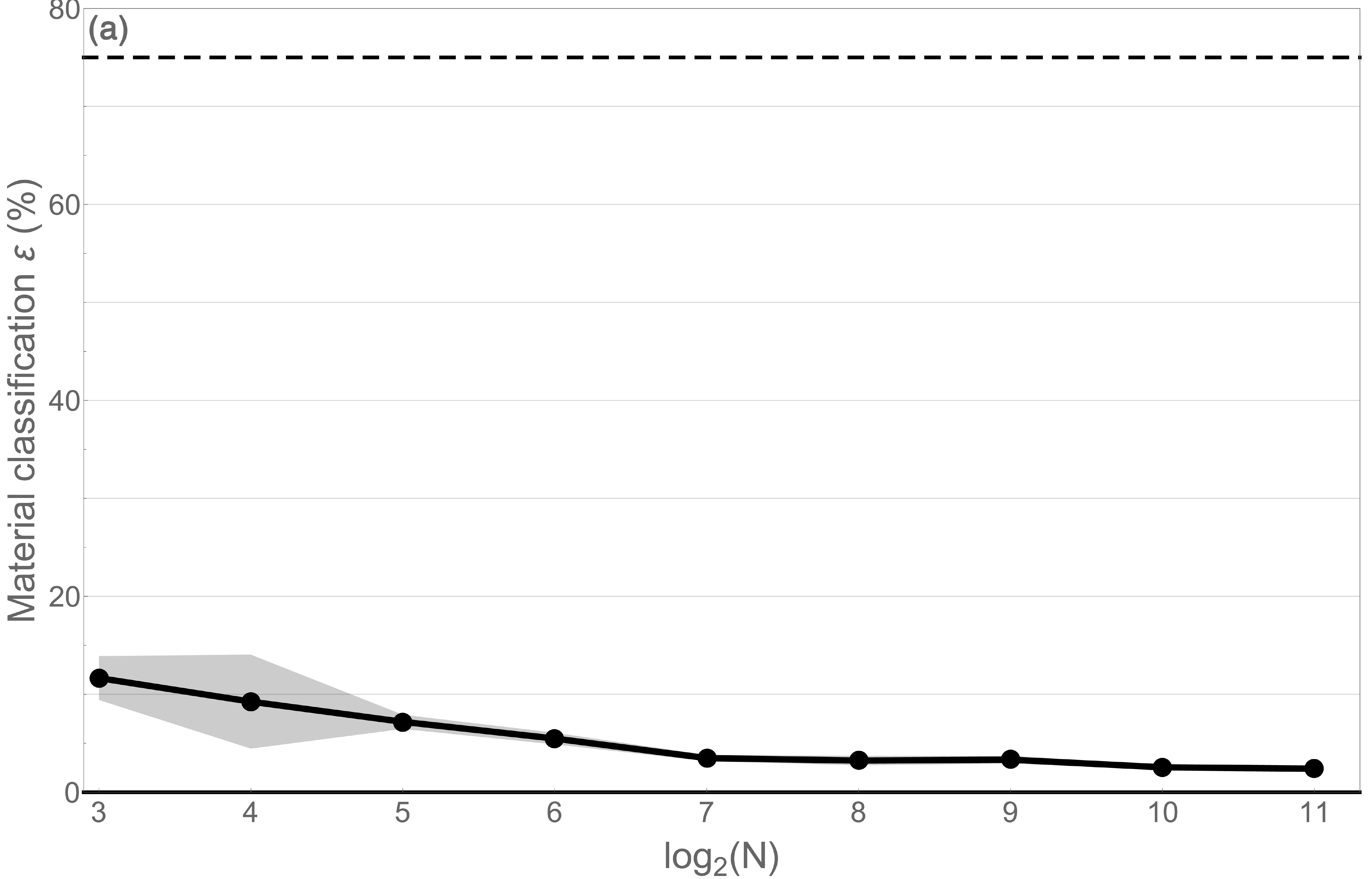}
\includegraphics[width=0.49\linewidth]{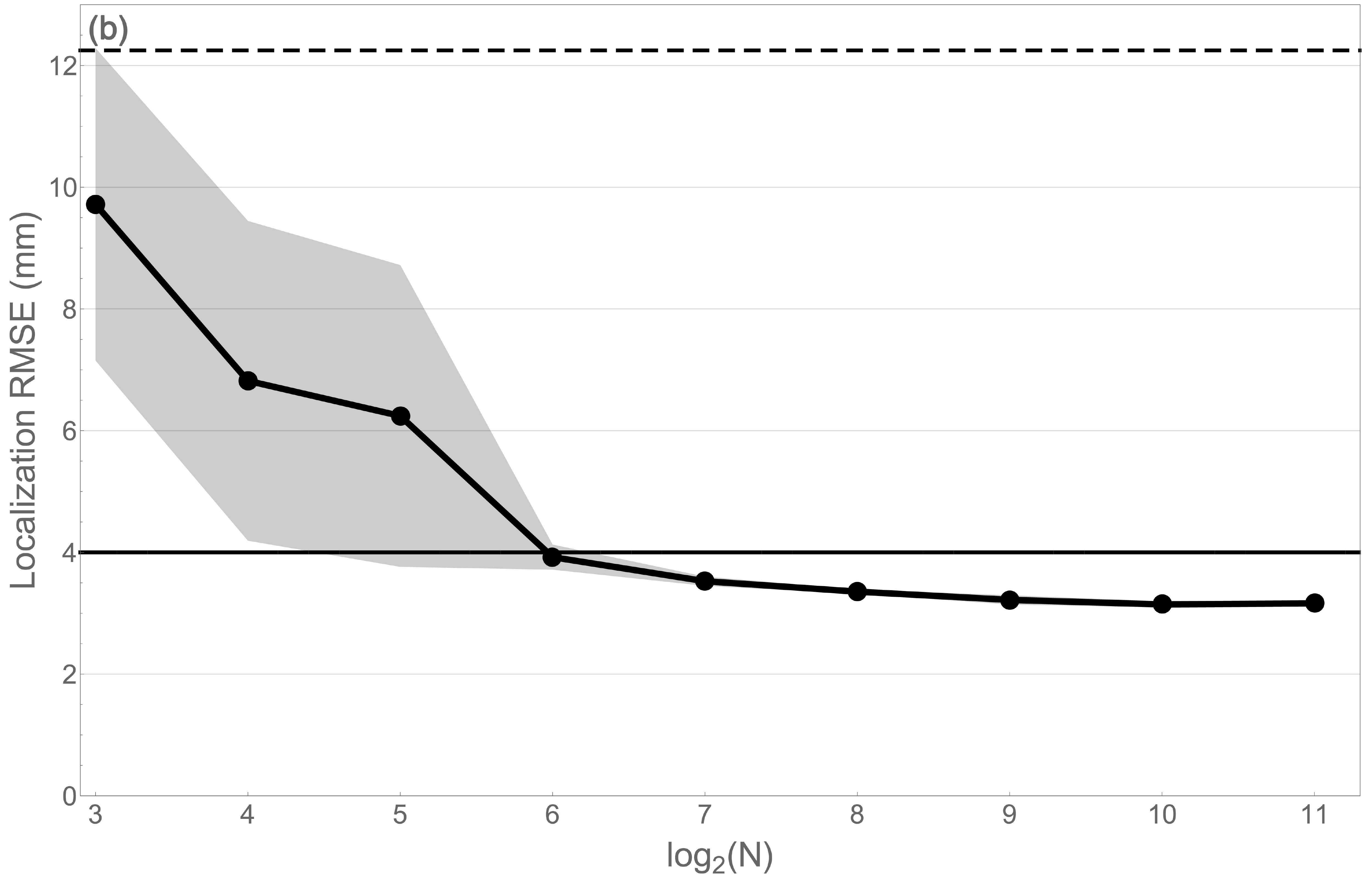}
\caption{\textbf{(a)} Material classification $\varepsilon$ {\lm as a function of N}. \textbf{(b)} Localization RMSE as a function of N. The shaded area marks the 95\% CI of the mean cross-validated performance. The dashed lines mark the baseline (chance) performance. The horizontal thick lines mark the ceiling performance.}\label{fig:materials}
\end{figure}

As shown in Fig.~\ref{fig:materials}, the four tested materials produced excellent results both in terms of classification and of localization. 

The confusion matrix reported in Tab.~\ref{tab:confusionmaterials} further confirms the success of the approach.

\begin{table}[htbp]
\begin{center}
\begin{tabular}{c|c|c|c|c|}
 & Cu & Al & Ti90/Al6/V4 & Graphite \\
\hline
\textit{Cu} &  \textbf{94.5\%} & 4.8\% & 0.0\% & 0.7\% \\
\hline
\textit{Al} & 4.9\%  & \textbf{95.0\%}  & 0.0\% &  0.1\%\\
\hline
\textit{Ti90/Al6/V4} & 0.0\% & 0.0\% & \textbf{99.9\%} & 0.1\% \\
\hline
\textit{Graphite} & 1.9\% & 1.2\% & 0.0\%  & \textbf{96.9\%}  \\
\hline
\end{tabular}
\caption{Materials classification confusion matrix based on N=256 training images. Rows are predictions.}\label{tab:confusionmaterials}
\end{center}
\end{table}

\begin{figure}[htbp]
\includegraphics[width=\linewidth]{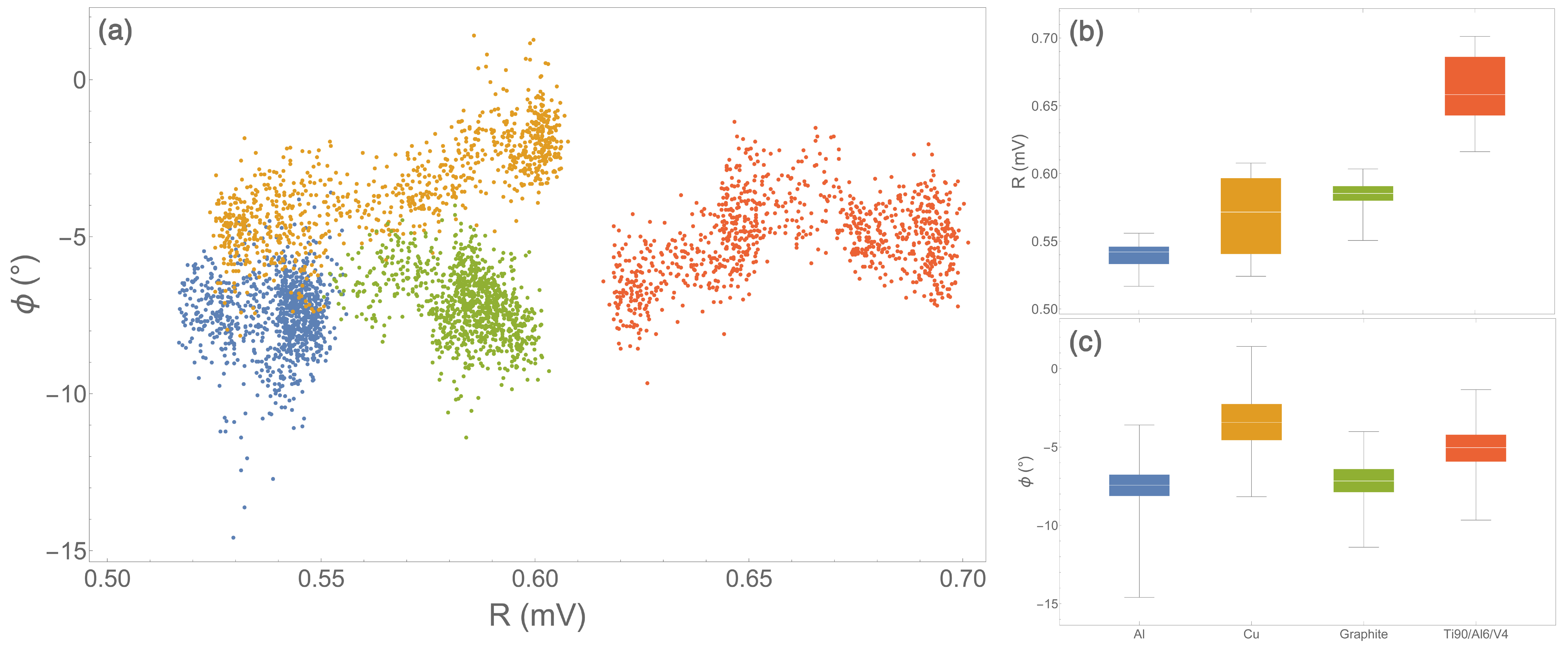}
\caption{\textbf{(a)} Scatter plot of the complete materials dataset: mean R and $\Phi$ of each image in the dataset. \textbf{(b)},  \textbf{(c)} Box plots of corresponding overall mean R and $\Phi$, respectively. The whiskers indicate the extreme values of the dataset, the thin white line the median of the dataset, the boxes mark interquartile range.}\label{fig:comparison}
\end{figure}

In each case, classification performed very well with a small incidence ($\leq$5\%) of confusion between single materials. The highest rates of incorrect classification are symmetric between Cu and Al. We attribute this fact to the similar EMI signals produced by the two materials, given their similar conductivities (see also Tab.~\ref{tab:materials}).  This is confirmed by Fig.~\ref{fig:comparison}(a), which shows the mean values of R and $\Phi$ of all the images of the dataset. By simultaneously using information from R and $\Phi$, populations are clearly distinguishable, with the exception of Cu and Al, which are partially overlapped in the scatter plot. This argument is further supported by the box plots of the corresponding mean values calculated across the entire dataset (Figs.~\ref{fig:comparison}(b),(c)). In summary, while the task of discriminating graphite and Ti90/Al6/V4 would have been reasonably easy also for a human observer, discrimination between Cu and Al (see also Fig.~\ref{fig:examples}(a),(e) and ~\ref{fig:examples}(b),(f)) would have been more challenging. Nevertheless, ML completes this tasks with a very high success rate.

Localization (Fig.~\ref{fig:materials}(b)) produced excellent results with any material. The ceiling performance RMSE$_{h}$=4~mm is exceeded for N$\geq$64. Further increase in the training dataset population reduced the RMSE down to 3.1~mm with N=2048.

\subsection{Shape dataset - Localization}
\begin{figure}[htbp]
\includegraphics[width=0.49\linewidth]{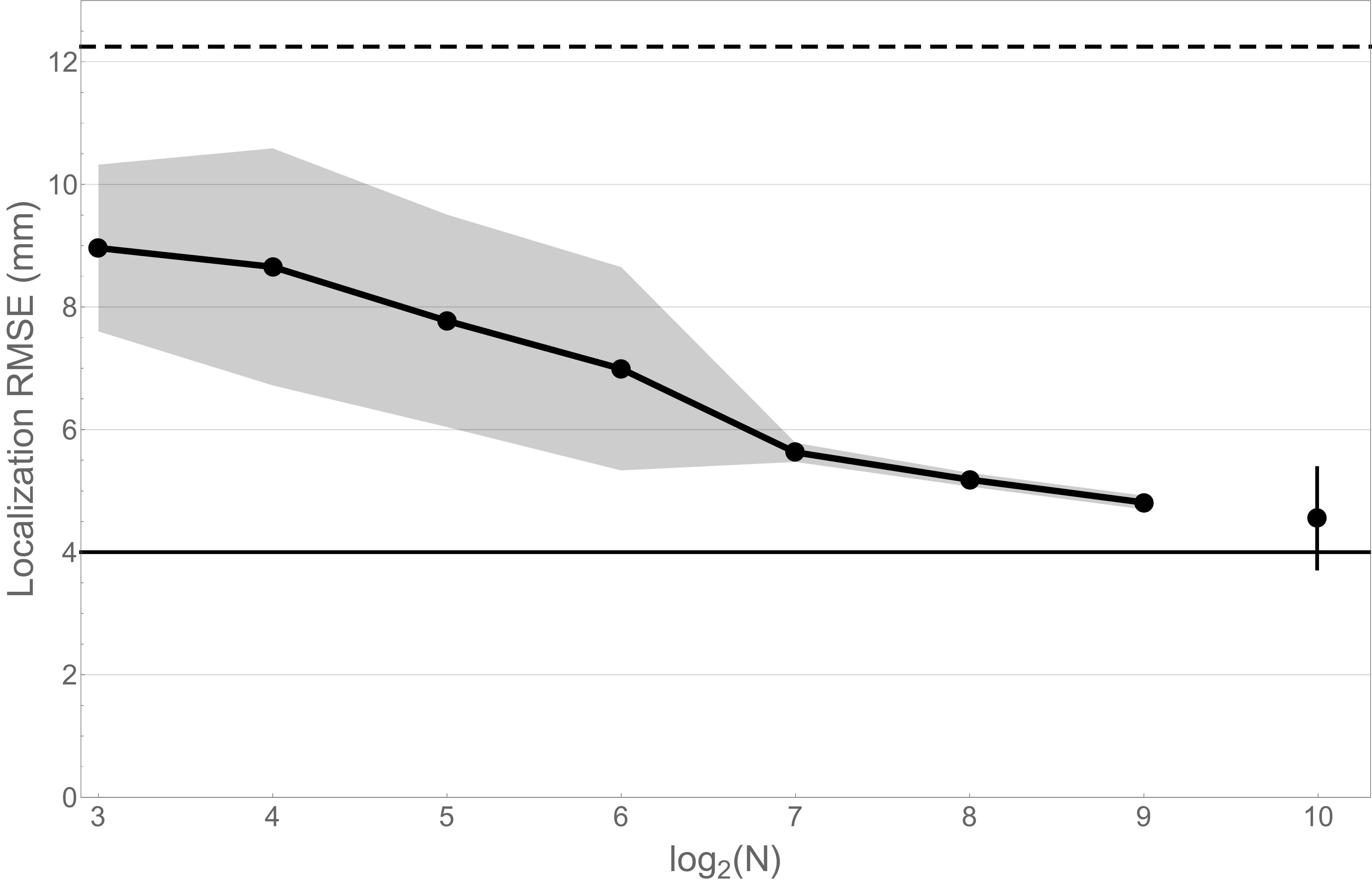}
\caption{Localization RMSE as a function of N for the shape dataset. The isolated marker represent the blind data results on a sample N=40. he shaded area marks the 95$\%$ confidence interval of the mean cross-validated performance. The error bars for the blinded results indicate the 95$\%$ CIs of system performance given the limited amount of blinded data. The dashed line marks the baseline (chance) performance. The horizontal thick line marks the ceiling performance.}\label{fig:shapeposition}
\end{figure}

The localization results for the shape dataset are shown in Fig.~\ref{fig:shapeposition}. As in the case of the sample B with 4~mm step (Fig.~3 in the main text), the blind dataset performed better than the cross-validation sets.

\end{document}